\documentclass[12pt]{iopart}

\usepackage{amssymb}
\usepackage{times}
\usepackage[dvipdf]{graphicx}
\usepackage{hyperref}

\begin{document}

\title[Generation of four-qubit cluster of entangled coherent states in
bimodal QED cavities]{Generation of four-qubit cluster of entangled coherent states in
bimodal QED cavities}

\author{E. M. Becerra-Castro, W. B. Cardoso$^{*}$, A. T. Avelar$^{\dagger}$ and B. Baseia}

\address{Instituto de F\'{\i}sica, Universidade Federal de Goi\'{a}s, 74.001-970, Goi%
\^{a}nia (GO), Brazil.}

\ead{$^{*}$wesleybcardoso@gmail.com}
\ead{$^{\dagger}$avelar@if.ufg.br}

\begin{abstract}
A recent work \cite{Munhoz_LA07} proposed a type of cluster entangled
coherent states and its generation. Here we present an alternative
experimental arrangement for its generation in bimodal QED cavities. The
scheme employs a single two-level atom that interacts dispersively with
cavity modes initially prepared in coherent states. The fidelity and success
probability of the state preparation are obtained considering the influence
of atomic velocity spread and atomic efficiency detection.

\end{abstract}

Entanglement is the most fantastic phenomenon of quantum mechanics and plays
a fundamental role in quantum information theory and applications. There are
various kinds of entangled states depending on the number of involved
parties, e.g.: for two qubits there is only one example, the EPR state \cite%
{Lo_PRA01}; for three qubits there are two kinds of three-partite
entangled states, the GHZ \cite{Dur_PRA00} and W \cite{Acin_PRL01}
states; for four-qubit states there are at least nine kinds of
entangled states, also
named cluster states. In particular, in Refs.\cite%
{Walther_PRL05,kielsen_PRL05} it was experimentally demonstrated that
correlations in a four-qubit, linear cluster states given by
\begin{equation}
|\phi _{4}\rangle =\frac{1}{2}\left( |0000\rangle +|0011\rangle
+|1100\rangle -|1111\rangle \right)
\end{equation}%
cannot be described in terms of the local realism. In fact, this state is
not biseparable and has a genuine four-qubit entanglement \cite%
{Tokunaga_PRA06}. Recently, Walther \textit{et al.}
\cite{Walther_NAT05} have implemented four-qubit cluster states
encoded into the polarization state of four photons, which
constituted an experimental proof of one-way quantum computing in
Raussendorf and Briegel's "one-way" model \cite{Raussendorf_PRL01}
using initialized qubits in a highly entangled cluster state. In a
recent paper Blythe and Varcoe \cite{BlytheNJP06}, using the
techniques of cluster state quantum computing, they showed how a
scalable quantum computer could potentially be constructed using
microwave cavity quantum electrodynamics (QED). These cluster state
quantum computing uses a sequence of measurements on a lattice of
entangled qubits in cluster states to perform quantum gate
operations.

Based on potential applications of cluster states, schemes for their
generation have been presented in various scenarios \cite%
{Munhoz_LA07,kielsen_PRL05,Walther_NAT05,Zou_PRA04,Zou_PRA05,
Zhou_PRA05-2,Barrett_PRA05,Borhani_PRA05,Cho_PRL05,Tokunaga_PRA05}.
In \cite{Zou_PRA04} Zou and Mathis it was proposed an unified scheme
to generate GHZ states, W states, and cluster states of four distant
atoms that are trapped separately in leaky cavities. They have also
studied: a scheme to generate a four-photon polarization-entangled
cluster state using only linear optical elements \cite{Zou_PRA05}
and four-photon coincidence detection; and two additional schemes to
generate the cluster states in the context of QED cavity
\cite{Zhou_PRA05-2}. Barrett and Kok \cite{Barrett_PRA05} proposed a
protocol to generate cluster states using spatially separated matter
qubits and single-photon interference effects
\cite{Borhani_PRA05,Cho_PRL05}. In Ref.\cite{Tokunaga_PRA05} the
authors proposed methods of fidelity estimation and entanglement
verification to experimentally produce four-qubit cluster states.

Recently, Munhoz \textit{et al.} \cite{Munhoz_LA07} investigated a
cluster-type of four-qubit coherent states and its generation\ using five
QED cavities, two Ramsey zones, a pair of two-level Rydberg atoms, an
external classical field plus atomic ionization detectors. In view of the
interesting idea and its potential applications, in this report we will
present an alternative and simplified scheme for its generation in the same
QED cavity context. We employ a pair of QED bimodal cavities, a single
two-level (Rydberg) atom and a dispersive atom-field interaction. The
simplicity of our scheme makes it attractive experimentally due to its
feasibility in the present status of QED technology \cite%
{Raimond01,Gleyzes_NAT07}. This proposal requires the experimental
setup shown in Fig. 1. The source \textit{S} ejects rubidium atoms
which are velocity selected and prepared in the circular Rydberg
state, one at a time, by appropriated laser beams in box the
\textit{B}. The relevant atomic levels $\left\vert g\right\rangle $
and $|e\rangle $, with the principal quantum numbers 50 and 51,
provide the atomic transition of $51.1$ $GHz$. \emph{R$_{1}$},
\emph{R$_{2}$} and \emph{R$_{3}$} stand for Ramsey zones which
perform a resonant $ \pi /2$ pulse on the $e\rightarrow g$
transition. The two bimodal high-Q superconducting cavities $C_{1}$
and $C_{2}$ are Fabry-Perot resonators made of two spherical niobium
mirrors with two orthogonally polarized TEM$_{900}$
modes separated by $1.2$ $MHz$ having the same Gaussian geometry (waist $%
w=6~mm$), and photon damping times of $130~ms$ \cite{Gleyzes_NAT07}; $S$
stands for the microwave generator coupled to each cavity, $D_{e}$ and $%
D_{g} $ are atomic detectors. These cavities are prepared at low temperature
($T\simeq 0.6$ $K$) to reduce the average number of thermal photons; before
starting the experiment the thermal field is erased \cite{Nogues99}.

To generate the cluster states in the four mode of the two bimodal cavities
we need to make two operations: i) The first one occurs when the atom
crosses the Ramsey zones R$_{1}$, R$_{2}$, and R$_{3}$ and interacts with the%
\textbf{\ }classical fields resonant to the atomic transition between the
states $\left\vert e\right\rangle $ and $\left\vert g\right\rangle $, with
intensities adjusted to produce a $\pi /2$ rotation in the atomic space,
i.e.,
\begin{eqnarray}
\left\vert e\right\rangle &\longrightarrow &(\left\vert g\right\rangle
+\left\vert e\right\rangle )/\sqrt{2}\,,  \label{zre} \\
\left\vert g\right\rangle &\longrightarrow &(\left\vert g\right\rangle
-\left\vert e\right\rangle )/\sqrt{2}\,;  \label{zrg}
\end{eqnarray}%
ii) the second operation occurs in $C_{1}$ and $C_{2}$ and involves
dispersive atom-field interactions with these cavity modes, one at a time,
in such a way that the atom crossing the cavity in the excited state $%
\left\vert e\right\rangle $ ($\left\vert g\right\rangle )$ produces a
negative (positive) phase-shift in the desired mode state. The dispersive
interaction is described by the effective atom-field Hamiltonian \cite%
{Raimond01},
\begin{equation}
H_{ef}=\hbar \frac{g_{\beta }^{2}}{\delta _{\beta }}\left[ (\hat{a}_{\beta
}^{\dagger }\hat{a}_{\beta }+1)|e\rangle \langle e|-\hat{a}_{\beta
}^{\dagger }\hat{a}_{\beta }|g\rangle \langle g|\right] ,  \label{disp}
\end{equation}%
where $\hat{a}_{\beta }$ ($\hat{a}_{\beta }^{\dagger }$) is the boson
annihilation (creation) operator for mode $\beta $, with $\beta =A,B$ for
the cavity $C_{1}$; the same is valid for the cavity $C_{2}$ with $\beta
=C,D $; $|e\rangle \langle e|$ and $|g\rangle \langle g|$ are the atomic
projectors, $g_{\beta }$ stands for the vacuum Rabi coupling with mode $%
\beta $, and $\delta _{\beta }$ is the detuning between atomic transition
and the mode $\beta $. Although the cavity has two modes, the frequency
splitting of $1.2$ $MHz$ between them ensures the atom being efficiently
coupled only to a single mode \cite{Gleyzes_NAT07}. Then, using the Stark
effect we choose the detuning with the mode $\beta $ which will interact in
dispersive regime, i.e., $g_{\beta }^{2}\bar{n}/\delta _{\beta }\ll 1$,
where $\bar{n}$ stands for the average photon number in the cavity. The
evolution operator associated with Eq.(\ref{disp}) reads
\begin{equation}
\hat{U}_{ef}=e^{-i\phi _{\beta }(\hat{a}_{\beta }^{\dagger }\hat{a}_{\beta
}+1)}|e\rangle \langle e|+e^{i\phi _{\beta }\hat{a}_{\beta }^{\dagger }\hat{a%
}_{\beta }}|g\rangle \langle g|,
\end{equation}%
where $\phi _{\beta }=g_{\beta }^{2}t_{\beta }/\delta _{\beta }$ and $%
t_{\beta }$ is the interaction time of the atom and the field in mode $\beta
$\emph{.}
\begin{figure}[t]
\begin{center}
\includegraphics[{width=9cm}]{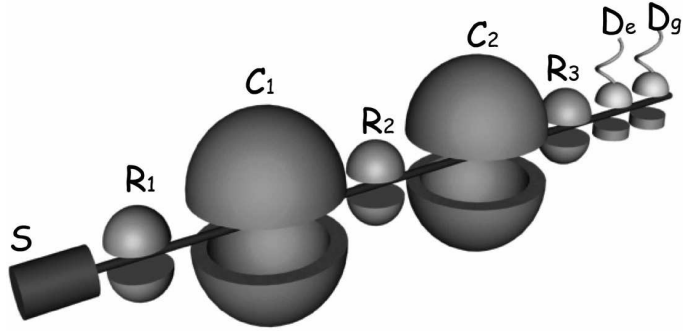} 
\end{center}
\caption{Schematic of the experimental setup for production of cluster states
in bimodal QED cavities.}
\label{cavidade}
\end{figure}

Now we discuss our procedure to generate the cluster coherent state
inside a two bimodal cavities. Initially, all the modes $A$ and $B$
of $C_{1}$ and $C$ and $D$ of $C_{2}$ are prepared in a coherent
state $\left\vert i\alpha \right\rangle $ whereas the atom is
prepared in the ground or excited states. The initial and final
atomic states determine the type of cluster state obtained. Before
entering in the first cavity the atom crosses a
Ramsey zone R$_{1}$ and the atomic state evolves to the superposition (\ref%
{zre}) and (\ref{zrg}). Thus, the state of the whole system can be written as%
\begin{equation}
\left\vert \psi \right\rangle _{R_{1}}=\frac{1}{\sqrt{2}}\left( \left\vert
g\right\rangle \pm \left\vert e\right\rangle \right) \left\vert i\alpha
\right\rangle _{A}\left\vert i\alpha \right\rangle _{B}\left\vert i\alpha
\right\rangle _{C}\left\vert i\alpha \right\rangle _{D},
\end{equation}%
where the sign $+$ ($-$) comes from the atom initially prepared in the
excited (ground) state. After that it enters the first cavity (detuning $%
\delta _{A}$) and interacts dispersively with the mode $A$ during a time $%
t_{A}$ to produce a phase shift $\phi _{A}=\pi /2$. At the cavity axis, a
Stark effect is applied to the atom tuning it dispersively ($\delta
_{B}=\delta _{A}$) with the mode $B$ for a time $t_{B}=t_{A}$.\ As result
the state of the whole system becomes
\begin{eqnarray}
\left\vert \psi \right\rangle _{C_{1}}& =&\frac{1}{\sqrt{2}}\left( \left\vert
g\right\rangle \left\vert -\alpha \right\rangle _{A}\left\vert -\alpha
\right\rangle _{B}\left\vert i\alpha \right\rangle _{C}\left\vert i\alpha
\right\rangle _{D} \mp \left\vert e\right\rangle \left\vert \alpha \right\rangle
_{A}\left\vert \alpha \right\rangle _{B}\left\vert i\alpha \right\rangle
_{C}\left\vert i\alpha \right\rangle _{D}\right) .
\end{eqnarray}%
When the atom emerges from the first cavity it crosses another Ramsey zone R$%
_{2}$, tuned to the transitions (\ref{zre}) and (\ref{zrg}). In this way we
obtain
\begin{eqnarray}
\left\vert \psi \right\rangle _{R_{2}}& =&\frac{1}{2}\left( \left\vert
g\right\rangle \left\vert -\alpha \right\rangle _{A}\left\vert -\alpha
\right\rangle _{B}\left\vert i\alpha \right\rangle _{C}\left\vert i\alpha
\right\rangle _{D} -\left\vert e\right\rangle \left\vert -\alpha \right\rangle _{A}\left\vert
-\alpha \right\rangle _{B}\left\vert i\alpha \right\rangle _{C}\left\vert
i\alpha \right\rangle _{D} \right.  \nonumber \\
& \mp & \left. \left\vert g\right\rangle \left\vert \alpha \right\rangle
_{A}\left\vert \alpha \right\rangle _{B}\left\vert i\alpha \right\rangle
_{C}\left\vert i\alpha \right\rangle _{D}  \mp \left\vert e\right\rangle \left\vert \alpha \right\rangle
_{A}\left\vert \alpha \right\rangle _{B}\left\vert i\alpha \right\rangle
_{C}\left\vert i\alpha \right\rangle _{D}\right) .
\end{eqnarray}%
Next the atom enters the cavity $C_{2}$ and interacts dispersively with the
modes $C$ and $D$, in the same way as found in $C_{1}.$ So the whole state
results%
\begin{eqnarray}
\left\vert \psi \right\rangle _{C_{2}}& =&\frac{1}{2}\left( \left\vert
g\right\rangle \left\vert -\alpha \right\rangle _{A}\left\vert -\alpha
\right\rangle _{B}\left\vert -\alpha \right\rangle _{C}\left\vert -\alpha
\right\rangle _{D} + \left\vert e\right\rangle \left\vert -\alpha \right\rangle _{A}\left\vert
-\alpha \right\rangle _{B}\left\vert \alpha \right\rangle _{C}\left\vert
\alpha \right\rangle _{D} \right. \nonumber \\
& \mp & \left. \left\vert g\right\rangle \left\vert \alpha \right\rangle
_{A}\left\vert \alpha \right\rangle _{B}\left\vert -\alpha \right\rangle
_{C}\left\vert -\alpha \right\rangle _{D}  \pm \left\vert e\right\rangle \left\vert \alpha \right\rangle
_{A}\left\vert \alpha \right\rangle _{B}\left\vert \alpha \right\rangle
_{C}\left\vert \alpha \right\rangle _{D}\right) .
\end{eqnarray}%
Now, the atom crosses another Ramsey zone R$_{3}$, leading the whole system
to the state%
\begin{eqnarray}
\left\vert \psi \right\rangle _{R_{3}}& =&\frac{1}{2\sqrt{2}}\left[
\left\vert g\right\rangle \left( \left\vert -\alpha \right\rangle
_{A}\left\vert -\alpha \right\rangle _{B}\left\vert -\alpha
\right\rangle _{C}\left\vert -\alpha \right\rangle _{D} + \left\vert
-\alpha \right\rangle _{A}\left\vert -\alpha \right\rangle
_{B}\left\vert \alpha \right\rangle _{C}\left\vert \alpha
\right\rangle _{D}
\right. \right. \nonumber \\
& \mp& \left\vert \alpha \right\rangle _{A}\left\vert \alpha \right\rangle
_{B}\left\vert -\alpha \right\rangle _{C}\left\vert -\alpha \right\rangle
_{D}  \pm \left. \left\vert \alpha \right\rangle _{A}\left\vert \alpha
\right\rangle _{B}\left\vert \alpha \right\rangle _{C}\left\vert \alpha
\right\rangle _{D}\right)  \nonumber \\
& +&\left\vert e\right\rangle \left( -\left\vert -\alpha
\right\rangle _{A}\left\vert -\alpha \right\rangle _{B}\left\vert
-\alpha \right\rangle _{C}\left\vert -\alpha \right\rangle
_{D}\right.  + \left\vert -\alpha \right\rangle _{A}\left\vert
-\alpha \right\rangle _{B}\left\vert \alpha \right\rangle
_{C}\left\vert \alpha \right\rangle _{D}
\nonumber \\
& \pm & \left. \left. \left\vert \alpha \right\rangle _{A}\left\vert \alpha \right\rangle
_{B}\left\vert -\alpha \right\rangle _{C}\left\vert -\alpha \right\rangle
_{D}  \pm \left\vert \alpha \right\rangle _{A}\left\vert \alpha
\right\rangle _{B}\left\vert \alpha \right\rangle _{C}\left\vert \alpha
\right\rangle _{D}\right) \right] .
\end{eqnarray}

Finally the atomic detection projects the state of the two cavities in one
of the following four-qubit coherent cluster states
\begin{eqnarray}
\left|\chi_{1}^{(e,e)}\right\rangle &=&\frac{1}{2}\left(
\left\vert \alpha\right\rangle\left\vert \alpha\right\rangle\left\vert
\alpha\right\rangle\left\vert \alpha\right\rangle
+\left\vert \alpha\right\rangle \left\vert \alpha\right\rangle
\left\vert -\alpha\right\rangle \left\vert -\alpha\right\rangle \right. \nonumber \\
& +& \left. \left\vert -\alpha\right\rangle \left\vert
-\alpha\right\rangle
\left\vert \alpha\right\rangle \left\vert \alpha\right\rangle
-\left\vert -\alpha\right\rangle \left\vert -\alpha\right\rangle
\left\vert -\alpha\right\rangle\left\vert -\alpha\right\rangle
\right), \label{CC1}
\end{eqnarray}
\begin{eqnarray}
\left\vert \chi_{2}^{\left( g,g\right)  }\right\rangle &=&\frac{1}{2}\left(
-\left\vert \alpha\right\rangle \left\vert \alpha\right\rangle
\left\vert \alpha\right\rangle \left\vert \alpha\right\rangle
+\left\vert \alpha\right\rangle \left\vert \alpha\right\rangle
\left\vert -\alpha\right\rangle \left\vert -\alpha\right\rangle \right. \nonumber \\
& +& \left. \left\vert -\alpha\right\rangle \left\vert -\alpha\right\rangle
\left\vert \alpha\right\rangle \left\vert \alpha\right\rangle
+\left\vert -\alpha\right\rangle \left\vert -\alpha\right\rangle
\left\vert -\alpha\right\rangle \left\vert -\alpha\right\rangle
\right), \label{CC2}
\end{eqnarray}
\begin{eqnarray}
\left\vert \chi_{3}^{\left(  g,e\right)  }\right\rangle &=&\frac{1}{2}\left(
\left\vert \alpha\right\rangle \left\vert \alpha\right\rangle
\left\vert \alpha\right\rangle \left\vert \alpha\right\rangle
+\left\vert \alpha\right\rangle \left\vert \alpha\right\rangle
\left\vert -\alpha\right\rangle \left\vert -\alpha\right\rangle \right. \nonumber \\
& -& \left. \left\vert -\alpha\right\rangle \left\vert -\alpha\right\rangle
\left\vert \alpha\right\rangle \left\vert \alpha\right\rangle
+\left\vert -\alpha\right\rangle \left\vert -\alpha\right\rangle
\left\vert -\alpha\right\rangle \left\vert -\alpha\right\rangle
\right)   , \label{CC3}
\end{eqnarray}
\begin{eqnarray}
\left\vert \chi_{4}^{\left(  e,g\right)  }\right\rangle &=&\frac{1}{2}\left(
\left\vert \alpha\right\rangle \left\vert \alpha\right\rangle
\left\vert \alpha\right\rangle \left\vert \alpha\right\rangle
-\left\vert \alpha\right\rangle \left\vert \alpha\right\rangle
\left\vert -\alpha\right\rangle \left\vert -\alpha\right\rangle \right. \nonumber \\
& +& \left. \left\vert -\alpha\right\rangle \left\vert -\alpha\right\rangle
\left\vert \alpha\right\rangle \left\vert \alpha\right\rangle
+\left\vert -\alpha\right\rangle \left\vert -\alpha\right\rangle
\left\vert -\alpha\right\rangle \left\vert -\alpha\right\rangle
\right)  , \label{CC4}
\end{eqnarray}
where we have omitted the subscripts of the modes; the superscripts in Eqs.(%
\ref{CC1})-(\ref{CC4}) indicate in which way the atomic state is
prepared/detected. Note that the cluster states $\left\vert \chi
_{1}^{\left( e,e\right) }\right\rangle $, $\left\vert \chi _{3}^{\left(
g,g\right) }\right\rangle $, $\left\vert \chi _{3}^{\left( g,e\right)
}\right\rangle $ and $\left\vert \chi _{4}^{\left( e,g\right) }\right\rangle
$ correspond to $|CLUSTER_{\alpha }^{+}\rangle $, $|CLUSTER_{\alpha
}^{-}\rangle $, $|C_{\alpha }^{+}\rangle $ and $|C_{\alpha }^{-}\rangle $,
respectively, found in \cite{Munhoz_LA07}. In the same way, other types of
cluster states appearing in \cite{Munhoz_LA07} can be obtained by convenient
choices of initial states. For example, from the initial state $\left\vert
i\alpha \right\rangle _{A}\left\vert i\alpha \right\rangle _{B}\left\vert
i\alpha \right\rangle _{C}\left\vert -i\alpha \right\rangle _{D}$ we obtain
the cluster state $\left\vert L_{\alpha }^{\pm }\right\rangle $ given in
\cite{Munhoz_LA07}.

Next, we calculate the total time $\tau $ spent for the generation
of the four-qubit cluster coherent states. First, note that $\alpha
$ must be chosen in a way that the cluster coherent states
(\ref{CC1}-\ref{CC4}) be orthogonal and this choice determines the
value of detuning $\delta $ satisfying the dispersive condition
$\delta _{\beta }\gg g_{\beta }\sqrt{ \bar{n}}$. Setting $\alpha =2$
leads to $\langle \alpha |-\alpha \rangle =e^{-2\alpha ^{2}}\simeq
10^{-3.5}\simeq 0$ and $\delta _{\beta }=40g_{\beta }$. Now, using
recent experimental data \cite{Gleyzes_NAT07} for the central Rabi
frequency $g_{\beta }\simeq 2\pi \times 51~kHz$ (with $\beta =A,B$)
one obtains the interaction time $t_{\beta }=\pi \delta _{\beta
}/2g_{\beta
}^{2}=196~\mu s$ to produce a phase shift $\pi /2$ in the state of mode $%
\beta $. This requires the atomic velocity $v=\sqrt{\pi }%
w/(t_{A}+t_{B})=27~m/s$ to get an effective atom-field interaction
time when taking into account the variation of the Rabi frequency
due to the atomic motion across the Gaussian cavity mode. The
selected atomic velocity belongs to the typical interval available
in laboratories $20-500~m/s$. The total time required for the
interaction of the atom with the four modes in the two cavities is
$\tau =0.784ms$. The scheme presented here can be mathematically
generalized for hight orders of particles in cluster coherent state
simply increasing more cavities in the experimental apparatus.
However one should take into account the total time to accomplish
the procedure and like this to count the decoherence effects in the
field state in the cavity, that in those cases can be important.

\begin{figure}[t]
\begin{center}
\includegraphics[{width=8cm}]{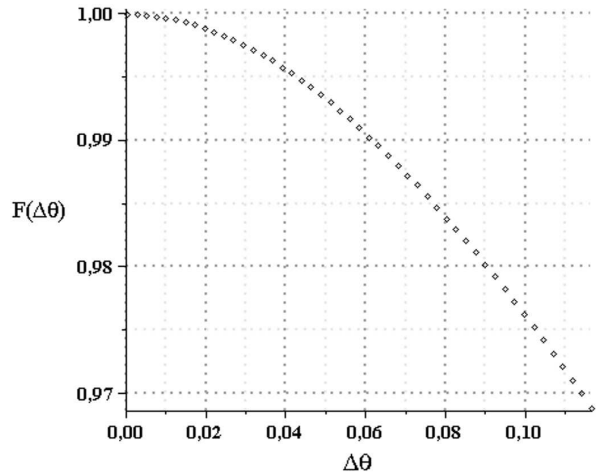} 
\end{center}
\caption{Fidelity of the prepared cluster state comparing the ideal
and real cases, obtained with the influence of the velocity spread
included.} \label{fid}
\end{figure}

Till now we have considered all calculations in an ideal experiment.
In a realistic case the effect of velocity spread, detection
efficiency, and dissipation should be taken into account. First, we
know that the atomic velocity determines the pulse $\theta $ in the
Ramsey zones and the phase shift $\phi $ in the cavity mode, so any
atomic velocity spread produces uncertainties in the values of
$\theta $ and $\phi $. We calculate the fidelity comparing the ideal
cluster and real cluster states, obtained with the influence of the
velocity spread included. For experimental velocity spreads $\Delta
v=$ $\pm 2$ m/s \cite{Raimond01} the fidelity is $\simeq97\%$ as
showed in Fig.(\ref{fid}); second, there is no problem with respect
to detector efficiency since the scheme works with a single atom;
third,  the total time $\tau $\ of experiment ($\tau =0.784$ $ms$,
neglecting the spatial separation between cavities) is much lesser
than photon damping times ($t_{d}=130~ms)$ and atomic radiative
times ($T_{a}$ $\simeq ~30$ $ms,$ for Rydberg atoms); so, we will
neglect dissipation and the scheme may be experimentally feasible in
the realm of microwave.

In summary, inspired on a previous work \cite{Munhoz_LA07}\ we have
presented an alternative scheme to generate a four-qubit cluster of
coherent states based on the potential applications for quantum
computing \cite{Raussendorf_PRL01,Walther_NAT05,BlytheNJP06}.Cavity
QED is widely believed to be an excellent system for quantum gate
operations \cite{BlytheNJP06}. The our method to generate the
cluster coherent state is suitable for applications in the scheme
for cluster-state quantum computing of the Ref. \cite{BlytheNJP06}.
It employs two bimodal QED cavities and a single two-level Rydberg
atom, constituting an economic version compared with
\cite{Munhoz_LA07}. Here we have considered the variation of the
Rabi frequency due to the atomic motion across the Gaussian cavity
mode: it results in the total time of experiment $\tau =0.784~ms$,
smaller than that in \cite{Munhoz_LA07}. In addition, the success
probability to get a specific cluster state of the family in
Eqs.(\ref{CC1})-(\ref{CC4}) is $50\%$, instead of the $25\%$ found
in \cite{Munhoz_LA07}. Finally, we have taken advantage of a recent
result by the Haroche's group \cite{Gleyzes_NAT07}, allowing us to
neglect the decoherence of the state during the generation process.
\vspace{-0.5cm}

\section*{Acknowledgments}

We thank the CAPES, CNPq, FUNAPE, Brazilian agencies, and PRPPG/UFG, for the
partial supports.


\section*{References}

\begin{thebibliography}{99}
\bibitem{Munhoz_LA07} Munhoz P P, Semi\~{a}o F L, Vidiella-Barranco A
and Roversi J A 2007 \textit{arXiv:0705.1549v1} [quant-ph]

\bibitem{Lo_PRA01} Lo H K and Popescu S 2001 \textit{Phys. Rev. A} \textbf{63}
022301

\bibitem{Dur_PRA00} Dur W, Vidal G and Cirac J I 2000 \textit{Phys. Rev. A} \textbf{%
62} 062314

\bibitem{Acin_PRL01} Acin A, Bru$\beta $ D, Lewenstein M and Sanpera A 2001
\textit{Phys. Rev. Lett.} \textbf{87} 040401

\bibitem{Walther_PRL05} Walther P, Aspelmeyer M, Resch K J and Zeilinger A 2005 \textit{Phys. Rev. Lett.} \textbf{95} 020403

\bibitem{kielsen_PRL05} Kiesel N, Schmid C, Weber U, T\'{o}th G, G\"{u}hne O, Ursin R and Weinfurter H 2005 \textit{Phys. Rev. Lett.} \textbf{95} 210502

\bibitem{Tokunaga_PRA06} Tokunaga Y, Yamamoto T, Koashi M and Imoto N 2006
\textit{Phys. Rev. A} \textbf{74} 020301(R)

\bibitem{Walther_NAT05} Walther P, Resch K J, Rudolph T, Schenck S, Weinfurter H, Vedral V, Aspelmeyer M and Zeilinger A 2005 \textit{Nature} \textbf{434} 169

\bibitem{Raussendorf_PRL01} Raussendorf R and Briegel H J 2001 \textit{Phys. Rev.
Lett.} \textbf{86} 5188

\bibitem{BlytheNJP06} Blythe P J and Varcoe B T H 2006 \emph{New J. Phys.}
\textbf{8} 231

\bibitem{Zou_PRA04} Zou X B, Pahlke K and Mathis W 2004 \textit{Phys. Rev. A}
\textbf{69} 052314

\bibitem{Zou_PRA05} Zou X B and Mathis W 2005 \textit{Phys. Rev. A} \textbf{72}
013809

\bibitem{Zhou_PRA05-2} Zou X B and Mathis W 2005 \textit{Phys. Rev. A} \textbf{71}
032308

\bibitem{Barrett_PRA05} Barrett S D and Kok P 2005 \textit{Phys. Rev. A} \textbf{71}
060310(R)

\bibitem{Borhani_PRA05} Borhani M and Loss D 2005 \textit{Phys. Rev. A} \textbf{71}
034308

\bibitem{Cho_PRL05} Cho J and Lee H W 2005 \textit{Phys. Rev. Lett.} \textbf{95}
160501

\bibitem{Tokunaga_PRA05} Tokunaga Y, Yamamoto T, Koashi M and Imoto N 2005
\textit{Phys. Rev. A} \textbf{71} 030301(R)

\bibitem{Raimond01} Raimond J M, Brune M and Haroche S 2001 \textit{Rev. Mod. Phys.}
\textbf{73} 565

\bibitem{Gleyzes_NAT07} Gleyzes S, Kuhr S, Guerlin C, Bernu J, Del%
\'{e}glise S, Hoff U B, Brune M, Raimond J M and Haroche S 2007 \textit{Nature}
\textbf{446} 297.

\bibitem{Nogues99} Nogues G, Rauschenbeutel A, Osnaghi S, Brune M, Raimond J M and Haroche S 1999 \textit{Nature} \textbf{400}
239
\end{thebibliography}
\end{document}